# Ultra-large tensile strains and martensite destabilization observed in high-temperature $Ni_{57.5}Mn_{22.5}Ga_{20.0}$ single crystal

V. A. Chernenko (1,2,#), E. Villa (3), V.A. Lvov (4), S. Besseghini (3), J.M. Barandiaran (1)

(1) *Universidad del País Vasco, Dpto. Electricidad y Electronica, P.O. Box 644, Bilbao 48080, Spain* , E-mail address: vladimir.chernenko@gmail.com
(2) *Ikerbasque, Basque Foundation for Science, Bilbao 48011, Spain*
(3) *CNR-IENI, C.Promessi Sposi, 29, Lecco 23900, Italy*
(4) *Institute of Magnetism, Vernadsky Str. 36-b, Kyiv 03142, Ukraine*

Tensile stress–strain behavior of $Ni_{57.5}Mn_{22.5}Ga_{20.0}$ single crystal exhibiting a high-temperature 2M-martensitic phase stable up to 360ºC has been studied in the course of thermal and mechanical cycling. The ultra-large reversible strains, about 9%, caused by the shape memory and superelasticity effects, have been observed up to 400 ºC being the instrumental temperature limit. Abnormally large two-way shape memory effect with 9% of strain magnitude has been found. The cycling procedure and the variation of thermal/mechanical routs of the training of samples revealed the destabilization (rejuvenation) of martensite. This physical effect is opposite to the well-known phenomenon of martensite stabilization. A destabilization effect is explained phenomenologically in terms of internal stressing of the alloy sample by the crystal defects.

# 1. Introduction

Off-stoichiometric Ni–Mn–Ga Heusler alloys are well-known for their excellent ability to be strongly deformed by small temperature change, weak uniaxial stress and magnetic field of moderate value. This feature is associated with the thermoelastic martensitic transformation (MT) and high mobility of twin boundaries in the martensitic phase [1–5]. The relevant physical effects are referred to as the ordinary shape memory, the superelasticity and the ferromagnetic shape memory (all of them being important for the engineering and medical applications).

The MT temperatures strongly depend on the composition of Ni–Mn–Ga alloys [3,6,7] through the electron concentration, characterized by the $e/a$ ratio [8]. This dependence triggered a tentative separation of the Ni–Mn–Ga alloys into three groups [3,8–10]. The alloys with high electron concentrations, $e/a > 7.7$, normally corresponding to compositions enriched by Ni and being relatively far from the stoichiometric $Ni_2MnGa$ compound, exhibit a martensitic transformation well above both room temperature and the Curie point. These alloys demonstrate a reversible MT from austenite to the non-modulated 2M-martensite above 200 ºC and attract an attention of researchers as promising high-temperature shape memory alloys (HTSMAs) [5–17]. Different aspects of these alloys have been already studied. Such studies mainly concerned the basic characteristics of MT in their interrelation with the alloy composition and microstructure [3,7–10,14] (including the addition of a forth element [18]), aging [15,17], and atomic ordering [16]. A number of reports about the compressive stress–strain tests of single crystals and polycrystalline Ni–Mn–Ga HTSMAs, showing improved plasticity, good shape recovery and thermal cycling stability, have been already presented [5,7,11,13,14,17]. The reported data point towards some advantages in the functional behaviour of these materials in comparison to the traditional Cu-, NiAl-, Zr-, NiTi(Zr,Hf)- and TiPd-based HTSMAs [19-22].

The interesting phenomenon, which is not well explored and comprehended yet, is a stabilisation of the non-modulated martensitic phase (2M-martensite) in the Heusler ferromagnetic shape memory alloys [12,15,23]. The non-modulated 2M-martensite is very attractive subject of inquiry because it corresponds to the ground state of these alloys [24] and shows a rather large magnetic field-induced strain of 0.2% [25]. Since the spontaneous lattice distortion during MT is almost as large as 20%, an actuation of huge strains is theoretically attainable under the different driving forces. This stimulates further studies of the high-temperature Ni–Mn–Ga alloys.

In the present work, ultra-large tensile deformations of the single crystals featuring shape memory (SM), two-way shape memory (TWSME) and superelastic (SE) effects in the high-temperature Ni–Mn–Ga HTSMA have been observed for the first time. We found that these remarkable functional abilities persist over the temperature range up to 400 °C being the temperature limit of our experimental equipment. The thermal/stress cycling through the temperature/stress range of MT has been systematically studied using different thermomechanical routes for the alloy treatment in order to clarify the effects of martensite stabilization/destabilization. These effects are explained in the framework of the model proposed recently for the description of martensite aging [26].

## 2. Experimental procedure

A single crystal, grown by the Bridgman method, with composition $Ni_{57.5}Mn_{22.5}Ga_{20.0}$ (in at.%, $e/a = 7.92$), as determined by X-ray fluorescent analysis, has been studied.

Calorimetric measurements have been carried out with a TA Q100 calorimeter at a temperature change rate of 10° C/min. The forward and reverse MT temperatures $T_m$ and $T_a$ were obtained, along with the transformation heat $q$, from the DSC curves as the respective peak positions and calculated areas under the peaks. The crystallographic orientation and the lattice parameters of the non-modulated body centred tetragonal unit cell of martensite, $a = 0.543$ nm, $c = 0.665$ nm, $c/a = 1.22$, were determined by using of PANalytical X'Pert Pro PW-3373/10 X'Celerator X-ray diffractometer.

A [100]-oriented sample measuring 0.57×0.19 mm$^2$ in a cross-section and 9.1 mm as a gauge length (sample S1 for the tensile tests) and the other sample with the dimensions of 0.28×0.27×8.5 mm$^3$ (sample S2) were spark cut from the single crystal. The samples were electropolished to remove the deformed surface layer. A TA Instruments Q800 dynamic mechanical analyzer (DMA) was used to measure (a) the tensile stress–strain (at constant temperature, $T_{exp}$) and strain–temperature curves in a quasi-static mode, and (b) the temperature dependence of the low-frequency elastic modulus, $E(T)$, in the dynamic tensile mode. The quasi-static and dynamic measurements were made using the same sample.

In the quasi-static mechanical tests, the upper limits of the load and the stress change rate were 18 N and 2.5 MPa/min, respectively. The rate of temperature change during the heating/cooling ramps of strain was equal to 10° C/min. The $E(T)$ dynamic measurements were carried out at a frequency of 1 Hz and oscillation strain amplitude of $10^{-4}$ with the rate of temperature change of 5° C/min. Further details of the dynamic method can be found elsewhere [27]. Note that in the DMA measuring unit, the thermocouple was not in direct contact with the sample, thus yielding a systematic shift between the temperature of the sample and the recorded one [27].

The cycling stability of MT temperatures was checked by DSC method. Afterwards, *two training routes* corresponding to the two samples were used to study shape memory and elastic properties.

*Route 1* was used to obtain a quasi-equilibrium stress–temperature phase diagram of the forward and reverse MTs versus temperature. To this end, after recording the $E(T)$ curve and holding at 400° C during 10 min, the sample S1 was subjected to stress–strain tests at a few constant temperatures $T_{exp}$. In this series of experiments, consisting in 15 stress–strain cycles, the values of $T_{exp}$ were lowered step-by-step from 400º C to room temperature. After each stress–strain cycle, the strain was recorded during the stress-free heating to 400º C, holding during 10 min and subsequent cooling to the next $T_{exp}$ value. At the end, $E(T)$ measurement was repeated.

*Route 2* was chosen to investigate the influence of the cyclic stress–strain training on the superelastic and thermal shape recovery properties of sample S2. The whole experiment consisted in a few series of 10 stress–strain cycles performed at fixed temperature values in the range from 400º C to $T_a$ with intermediate annealing at 400º C during 10 min after each series. After the first series of cycling (performed at 400º C) the unloaded sample S2 was subjected to 10 heating–cooling runs through MT.

## 3. Results

### 3.1. Stress-temperature phase diagram

The calorimetric measurements show the forward and reverse MT taking place in the temperature range from 300 ºC to 360 ºC. No other MTs have been observed between 20 ºC and 400 ºC. A detailed behaviour of DSC peaks during the 2$^{nd}$ and 11$^{th}$ cooling–heating ramps is depicted in Fig. 1. The Inset shows that the characteristic temperatures of the forward and reverse MTs decrease monotonically as a result of cycling but with slightly different rates, whereas the transformation heat tends to fluctuate near the value of 10 J/g.

The typical tensile stress–strain curves and the complementary zero-stress heating–cooling temperature dependencies of strain accumulated after each cycle of stressing for the sample S1 are shown in Figs. 2 and 3, respectively. At the temperatures below the forward MT temperature range, the sample deforms mainly due to the redistribution of martensite variants, while above this range a stress-induced MT occurs. Figs. 2 and 3 provide evidence of the remarkable superelastic, mechanical and thermomechanical characteristics of the Ni$_{57.5}$Mn$_{22.5}$Ga$_{20.0}$ single crystal, which were not observed before for the martensitic Heusler alloys at such high temperatures. First, the curves in Fig. 2 reveal a record-breaking (for high-temperature Ni–Mn–Ga alloys) value of the reversible tensile strain of about 9 % persisting up to 400 ºC, which is the instrumentally limited temperature. Second, a nearly 9% of pseudo-plastic deformation of the martensite caused by the fairly low detwinning stress is observed in the wide temperature range from 20ºC to 300 ºC. Third, Fig. 3 demonstrates a perfect thermal strain recovery (of about 9%) due to an ordinary shape memory effect, which manifests itself in the enormous sample contraction during the reverse MT. Finally, a rather large spontaneous elongation of the sample during the forward MT (of about of 4%) is observed as a manifestation of a two-way SME, which is hardly observed for the single crystalline samples of SMAs, as a rule. The aforementioned findings may be considered as a first step forward in the achievement of the high-performance functionality of high-temperature Ni–Mn–Ga alloys, in view of the extremely high values of theoretical limits of superelastic and pseudoplastic tensile deformations of these alloys, roughly estimated as 13% and 20%, respectively.

The martensite start and austenite finish stresses ($\sigma_{ms}$ and $\sigma_{af}$, respectively), which characterize the forward and reverse martensitic transformations of sample S1, were determined from stress–strain curves by a tangential method. These stresses are plotted in Fig. 4 as the functions of temperature. The data points are approximated by the straight solid lines with the slopes shown in Fig. 4. The third line in Fig. 4 demonstrates a relatively weak temperature dependence of the detwinning stress, $\sigma_{tw}$, of martensite observed in an unusually wide temperature range. The slope of the quasi-equilibrium stress–temperature phase diagram can be expressed by a generalized Clausius-Clapeyron relationship

$$\frac{d\sigma}{dT} = \frac{q\rho}{\varepsilon T_0}, \qquad (1)$$

where $q \approx 10$ J/g is the latent heat of MT (see Fig. 1), $\rho = 8 g/cm^3$ is the mass density, $T_0 = 598$ K is the temperature of equilibrium between the two phases (Fig.1), and $\varepsilon \approx 8\%$ is the elongation of the specimen in the direction of the applied stress (Fig.2). The calculated $d\sigma/dT$ value is equal to 1.8 MPa/ ºC, in agreement with the experimental values obtained by the linear approximation shown in Fig. 4.

Fig. 5 shows heating–cooling variations of *E*-modulus of the sample S1, which were measured during the first thermal cycle (curve 1) and after all the tests of Route 1 (curve 2). Both curves exhibit abrupt anomalies that are relevant to MT. These curves also provide an experimental evidence of the abnormally low value of *E*-modulus (≈ 10 GPa) in the austenitic phase. In the martensitic state, *E*-modulus increases on cooling of the sample. It is noteworthy that this modulus is about two times smaller after the thermomechanical cycling than before it. The other important effect of thermomechanical cycling is the shift of the MT hysteresis loop towards low temperatures.

Note, that the training procedure according to Route 1 does not lead to a martensitic stabilization effect (Fig. 3), that is typical e.g. for the Co–Ni–Ga and Ni–Fe(Co)–Ga single crystals [28,29]. On the contrary, it brings about an opposite effect, hereafter termed as the martensite destabilization (rejuvenation). Both training routes used in the present work reveal that the destabilization of martensite manifests itself in the reduction of the effective elastic modulus of martensite, in the decrease of transformation temperatures of MT (Figs. 1 and 5) and in the increase of critical stress values (see Sec. 3.2). This intriguing phenomenon is discussed in Sec. 4.

*3.2. Stress-strain training*

In this section, we demonstrate that the effects of martensite destabilization are much more pronounced after the stress-strain cycling procedure of Route 2, than in the case of Route 1, where this type of training was not performed, or during simple thermal cycling through MT. Typical results of $\sigma$-$\varepsilon$ cycling of sample S2 at two different temperatures are shown in Figs. 6 and 7. A series of 10 superelastic curves in Fig. 6 demonstrates a strong upward shift during cycling reflecting the increase of $\sigma_{ms}$. The reversible tensile deformation of about 10%, observed at 400°C, is the largest value known for shape memory alloys at such high temperature. Whereas the first few $\sigma$-$\varepsilon$ cycles yield one-step plateau, in the next cycles the transformation is characterized by several steps which, at the end, prevent to observe the full transformation due to the instrumental limitations. The small residual strain after each cycle may be explained by the presence of some amount of untransformed martensitic phase. This suggestion is supported by the results shown in the Inset to Fig.6, from which one can see that the thermal cycling of the trained sample S2 through MT is accompanied by the same reversible deformation of 10 % as in stress-strain tests. Such a huge spontaneous shape change at so high temperatures was not observed before. The sample S2, similarly to sample S1, shows a perfect shape recovery, contracting by 9% during heating through reverse MT, and exhibits the same magnitude of the extra-large spontaneous elongation during cooling through the forward MT due to TWSME, whereas S1 shows only a partial TWSME strain (see Fig.3). Thus, the appearance of the giant TWSME is the most prominent consequence of training Route 2. Furthermore, we found that training Route 2 leads to the practically important reproducible $\sigma$-$\varepsilon$ behavior which is observed only at temperatures close to $T_{ms}$ (Fig.7). Fig. 7 shows that this behavior is characterized by a stable value of $\sigma_{ms}$ and the disappearance of the flat plateau. Finally, $\sigma_{ms}$ values observed during cycling along Route 2 are plotted as a function of cycle number in Fig.8. The temperature dependence of the curves in Fig.8 will be discussed in the Sec. 4.

## 4. Discussion

A thermodynamic interpretation of the observed effects of the thermal/mechanical cycling through the temperature/stress range of MT can be made. To this end, the real crystal may be considered as a combined thermodynamic system consisting of the regular crystal lattice and crystal defects. The thermally or mechanically induced MT is accompanied by a slow reconfiguration of the defect subsystem, which drives the combined system to the equilibrium (see Ref. [30] and references therein). This process goes on during the martensite aging and results in the elevation of the reverse MT temperature as a signature of the martensite stabilization. Furthermore, the defect reconfiguration is accompanied by a slow variation of the average values of lattice parameters [31], which can be described by the slow-variable strain tensor components $\widetilde{\varepsilon}_{ik}(t)$. Recently, a stabilizing internal stress $\sigma_{ik}^{(s)}(t)$ was introduced and defined as a thermodynamic parameter which is conjugated to $\widetilde{\varepsilon}_{ik}(t)$ [26]. Hereafter we show how the concept of internal stress helps explaining the influence of the thermal and mechanical cycling on the shape-memory alloy behavior.

*i) Effect of thermal cycling.* Let the specimen be in the martensitic state at the initial moment of time $(t = 0)$, and let its thermal cycling start in the moment $t = t_0$. The thermal cycling of the specimen results in the reduction of the MT temperature, so it destabilizes the martensite. This martensite destabilization may be caused by the destabilizing internal stress defined as

$$\sigma_{ik}^{(d)}(n) \equiv \sigma_{ik}^{(s)}(t_0) - \sigma_{ik}^{(s)}(t_n), \tag{2}$$

where $t_n$ is the moment of the $n^{th}$ cycle performance. (The cycle duration is presumed to be small in comparison with the $t_0$ value).

A thermal cycling affects the combined system isotropically without changing the shape of the sample, and therefore, it induces only a destabilizing pressure $p^{(d)} = -\sigma_{ii}^{(d)}$. The destabilizing pressure (DP) can be obviously expressed as

$$p_a^{(d)}(n) = \left(\frac{dT_{ms}}{dp}\right)^{-1} \Delta T_{ms}(n), \quad (3)$$

for the austenitic phase, and

$$p_m^{(d)}(n) = \left(\frac{dT_{as}}{dp}\right)^{-1} \Delta T_{as}(n), \quad (4)$$

for the martensitic state. The values $\Delta T_{ms}(n) = T_{ms}(n) - T_{ms}(0)$ and $\Delta T_{as}(n) = T_{as}(n) - T_{as}(0)$ are the shifts of martensite start and austenite start temperatures observed after $n$ cycles. These temperature shifts are different in view of the obvious difference between the defect subsystems inherent to the austenitic and martensitic phases.

The derivatives involved in the Eqs. (3) and (4) can be evaluated from experiments on hydrostatic compression of the shape memory alloys. These values are positive [32], thus the destabilizing pressures are negative. It means that the thermal cycling leads to the increase of the specific volume of sample. The relative volume changes of austenite and martensite are expressed as

$$\frac{V_a(n) - V_a(0)}{V_a(0)} \stackrel{def}{\equiv} \frac{\Delta V_a(n)}{V} = -\frac{p_a^{(d)}(n)}{B_a}, \quad (5)$$

and

$$\frac{V_m(n) - V_m(0)}{V_m(0)} \stackrel{def}{\equiv} \frac{\Delta V_m(n)}{V} = -\frac{p_m^{(d)}(n)}{B_m}, \quad (6)$$

where $B_a$ and $B_m$ are the bulk moduli of austenitic and martensitic phases, respectively.

*ii) Effect of mechanical cycling.* Let the sample be a prism strongly elongated in $z$-direction with $z$-axis parallel to [001] crystallographic direction. In this case, an axial loading of the austenite sample in the [001] direction creates the stress

$$\sigma_{ik} = -p\delta_{ik} + \sigma_{zz}. \quad (7)$$

For the sake of certainty, we assume that the sign of this stress component coincides with the sign of the lattice distortion ($c/a - 1$), where $c$ and $a$ are the lattice parameters of tetragonal martensite. In this case a cyclic loading produces the $z$-variant of martensite during the stress-induced MT. A misfit between the crystal lattice and defect network configuration causes the internal stresses

$$\sigma_{a,ik} = -p_a^{(d)}\delta_{ik} + \sigma_{a,zz}^{(s)}, \tag{8}$$

and

$$\sigma_{m,ik} = -p_m^{(d)}\delta_{ik} + \sigma_{m,zz}^{(s)} \tag{9}$$

in the austenitic and martensitic phases, respectively.

The structure of the tensors represented by Eqs. (8) and (9) may be ascertained phenomenologically. The volume fractions of martensite variants formed on cooling the sample at $t=0$ are presumed to be equal to each other, and therefore, both the initial martensitic state and that aged during the period of time $t_0$, are isotropic. It means that the tensor $\sigma_{ik}^{(s)}(t_0)$ has only diagonal components, which are equal to each other. The mechanical cycling disturbs an equilibrium state of the crystal with defects, and so, creates the destabilizing pressures $p_m^{(d)} < 0$, $p_a^{(d)} < 0$. However, an adjustment of the defect subsystem to the lattice of $z$-variant of martensite goes on during the half-periods of cycling when a sample is in the martensitic state. This gives rise to the stabilizing axial stress (SAS) denoted as $\sigma_{m,zz}^{(s)}$.

*iii) The shifts of characteristic temperatures.* A thermal cycling induces a negative internal pressure, destabilizes the martensitic phase and reduces the characteristic temperatures of the forward and reverse MTs. The decrease of the reverse MT temperature may be more pronounced than that of the forward MT because the martensitic state is more defected, and therefore, $|\sigma_{m,ik}| > |\sigma_{a,ik}|$.

A mechanical cycling induces both destabilizing pressure and stabilizing axial stress. The DP uniformly expands all martensite variants. The SAS stabilizes the variant formed in the course of stress-induced MT and destabilizes the other variants. As so, the mechanical cycling "splits" the transformation temperatures of different variants. The scheme in Fig. 9 shows the effect of internal stressing on the martensite start and austenite start temperatures. For the sake of clarity, the shifting of the characteristic temperatures is conventionally subdivided into two stages: the DP reduces both temperatures at the Stage 1 and the SAS splits these temperatures at the Stage 2. The forward MT starts when the temperature of the cooled alloy reaches the highest of the splitted temperature values, whereas the reverse MT starts when the temperature of the heated alloy approaches the lowest of the splitted values.

The aforementioned considerations and Fig. 9 capture qualitatively the main features observed in the DSC measurements (Fig.1):

a) the effect of thermal cycling on the reverse MT temperature is more pronounced;
b) the temperature interval of the reverse MT becomes wider after cycling, due to the difference in the MT temperatures of different martensite variants;

c) the cycling reduces the temperature hysteresis of MT:

$$T_{as}(n) - T_{ms}(n) < T_{as}(0) - T_{ms}(0). \tag{10}$$

The only experimental feature, which is not grasped by the presented phenomenological considerations, is a slight narrowing of the temperature interval of the forward MT.

The DP and SAS concepts enable understanding the effects caused by mechanical cycling (Figs. 6 and 7). Accordingly, an upward shift of the stress–strain curves is caused by the cooperative action of DP and SAS. Note that a slight decrease of critical stress after the first cycle may be tentatively explained if we assume that the some SAS appears after the first cycle, while the creation of the sufficiently large DP needs more cycles. A large reversible deformation observed in the unloaded specimen after its mechanical cycling (TWSME) shows that SAS exceeds the twinning stress.

*iv) Quantitative estimations.* Eqs. (3)–(6) show that the DP value is proportional to the pressure shift of MT temperature, which, in turn, is directly proportional to the volume change $(\Delta V/V)_{MT}$ caused by MT [33]. The SAS may be assumed to be roughly proportional to the MT strain $(c-a)/a_0$. Note, that the relevant DP and SAS values are strongly different for the Ni–Mn–Ga martensites with $c/a < 1$ and $c/a > 1$. Table 1 shows the typical values of the appropriate physical parameters for these groups of martensites. The alloy studied in the present work belongs to the group with $c/a > 1$.

Table 1. Typical values of physical parameters characterizing the MTs of Ni–Mn–Ga alloys and their martensitic phases with $c/a > 1$ and $c/a < 1$ (see text for details).

|  | $c/a > 1$ | $c/a < 1$ |
|---|---|---|
| $(c-a)/a_0$ | 0.20 | –0.06 [33] |
| $dT_{ms}/dp$ (°C /GPa) | 25 [36] | 5.5 [33] |
| $(\Delta V/V)_{MT}$ | 0.005 [36] | 0.001 [33] |
| Twinning stress (MPa) | ~ 10[*] | ~ 1 [1] |

[*] See Fig. 4

Generally, the data of the Table 1 enable an estimation of the DP value from Eqs. (3) and (4) when the characteristic temperatures shifts caused by a thermal cycling of the sample are known. Incidentally, the experimental results presented in Fig.1 are not convenient for the precise evaluation of the temperature shifts involved in the Eqs. (3) and (4), especially for the forward MT. More pronounced are the DMA data in Fig.5 which show that the cycling of the sample S1 shifts the MT temperature range by a value of about –25 °C. This shift results in a DP value of –1.0 GPa, bringing in a volume change of about $\Delta V/V \approx 6\times10^{-3} \approx (\Delta V/V)_{MT}$ caused by the thermal cycling (see also Table 1). In the latter estimation, the value of the bulk modulus $B = (C_{11} + 2C_{12})/3 \approx 150$ GPa [34, 35] was used.

A shift of the martensite start temperature due to the mechanical cycling can be roughly estimated from the formula

$$\Delta T_{ms} \approx \left(\frac{d\sigma_{ms}}{dT}\right)^{-1} \Delta\sigma_{ms} \qquad (11)$$

The experimental value of the derivative involved in the Eq. (11) is 1.4 MPa/°C (Fig. 4). This value may be used for estimating the martensite start temperature shifts (see Table 2) from the experimentally observed changes of the critical stress after 10 stress–strain superelastic cycles (see Fig. 8).

Table 2. Influence of the 10 superelastic cycles on experimental shift of critical stress and corresponding calculated MT temperature shift

| $T_{exp}$ (°C) | 330 | 360 | 400 |
|---|---|---|---|
| $\Delta\sigma_{ms}$(MPa) | 1 | 30 | 59 |
| $\Delta T_{ms}$(°C) | –0.7 | –21 | –42 |

As follows from Table 2, both the observed change of the critical stress and an absolute value of the estimated change of the martensite start temperature quickly decrease on cooling the alloy. This feature evidences the decrease of mobility of the crystal defects affected by the internal stress. It should be mentioned, however, that the low-temperature mechanical cycling was performed after the high-temperature one. In this case the recovery of the defect configuration disturbed by the high-temperature cycling could be incomplete and this detail also could diminish the effect of the next cycling. In-situ microscopic observations of the defect structures are necessary to clarify the before-discussed mechanisms of martensite destabilization in this alloy. Such observations were hindered in the present work by the small size of the samples and high values of temperatures.

## 5. Conclusions

In the present work a $Ni_{57.5}Mn_{22.5}Ga_{20.0}$ single crystal exhibiting martensitic transformation into 2M-martensitic phase at high temperature was extensively studied under tensile load. The experimental stress–temperature phase diagram was plotted. The slopes of the experimentally obtained lability lines of the martensitic and austenitic phases agree with the Clausius-Clapeyron relationship. Record-breaking values of reversible strains of about 9% have been observed in the temperature range from 300 °C to 400 °C due to superelasticity, ordinary shape memory, and two-way shape memory effects. We consider that the experimentally observed two-way shape memory effect and the MT temperature shifts are related to the destabilization (rejuvenation) of martensite caused by the thermal/mechanical training of the alloy samples. The thermal/tensile training procedures give rise to two opposite tendencies: (i) a destabilization of the martensitic structure as a whole due to the uniform dilatation of the crystal lattice by the isotropically introduced defects (resulting in a decrease of the MT temperatures) and (ii) a stabilization of the favorable martensitic variant and destabilization of the other ones due to their internal uniaxial stressing. Both experiment and theory confirm that the thermal cycling of the loaded or unloaded sample through the MT is more effective to assist the tendency (i), whereas the superelastic/pseudoplastic stress–cycling favors the tendency (ii). In comparison with the stable self-accommodated state of the virgin martensite, the rejuvenated martensite should be characterized by interdependent factors such as a non-equilibrium coarse microstructure, reduced elastic modulus, increased number of defects and large isotropic and/or anisotropic internal stresses.

## Acknowledgements

V.A. Lvov is grateful to UPV/EHU for financing his stay at the Department of Electricity and Electronics. The financial support from the Department of Education, Basque Government (Project No. IT-347-07) and the Spanish Ministry of Education and Science (Project No. MAT2008 06542-C04-02) is acknowledged.

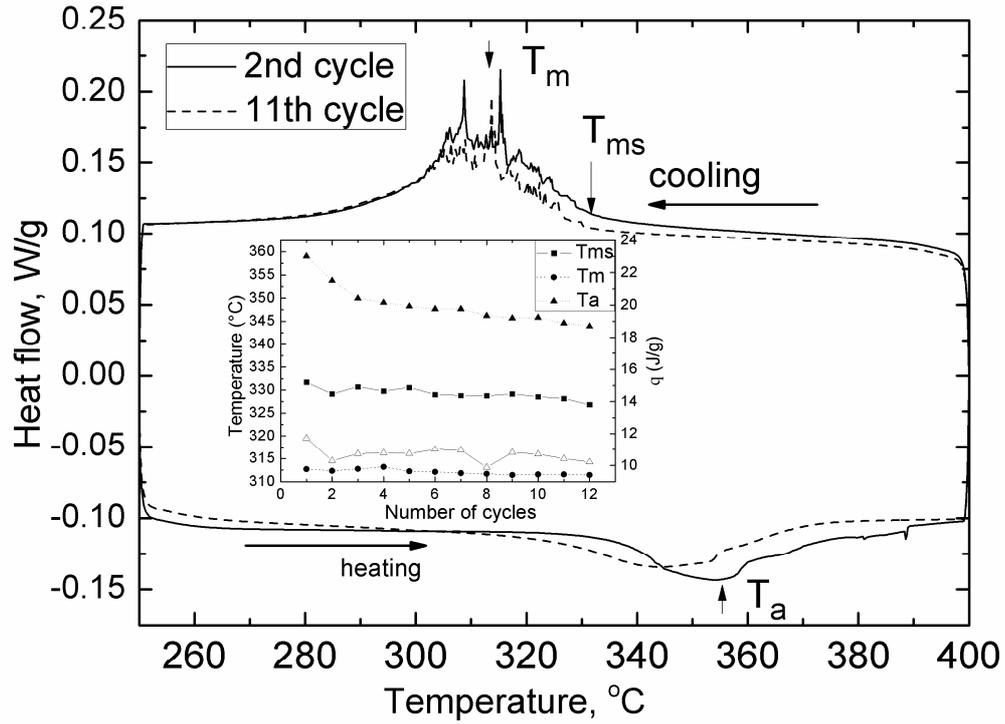

Fig.1. Heating-cooling DSC thermograms at the 2nd and 11th cycles. Inset: the MT temperature and transformation heat as a function of cycle number.

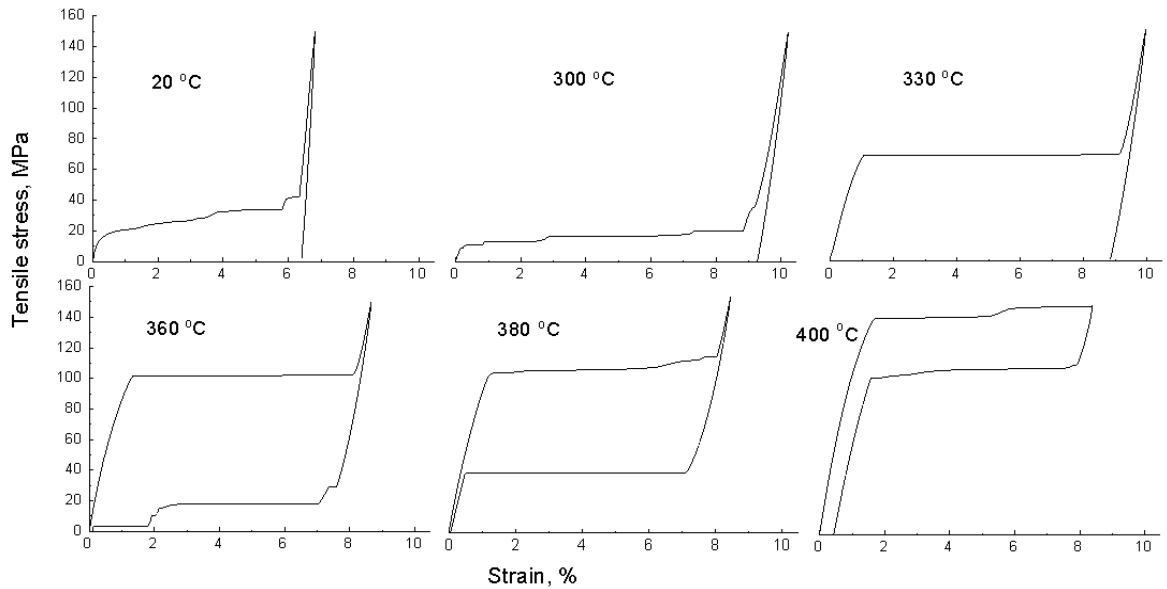

Fig.2. Selected tensile stress-strain curves of S1 single crystal along [100] axis showing ultra-large high-temperature superelasticity and large martensitic plasticity and low yield stress of martensite in a wide temperature range. The measurements are made in a sequence from 400°C to the room temperature (see text for details).

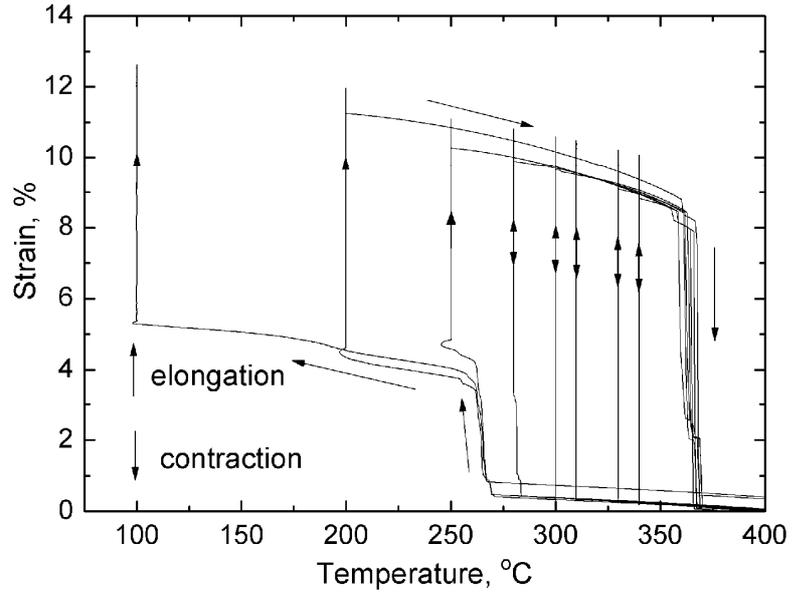

Fig.3. Selected stress-free heating-cooling strain dependencies showing perfect thermal recovery (contraction) and partial two-way SME (elongation) of the specimen S1. The curves were recorded during heating from $T_{exp}$ to 400°C after each $\sigma$-$\varepsilon$ test (Fig. 2) and subsequent cooling to the next temperature $T_{exp}$. The loading/unloading tracks of $\sigma$-$\varepsilon$ tests are shown as the vertical lines at the positions on $x$ axis equal to $T_{exp}$.

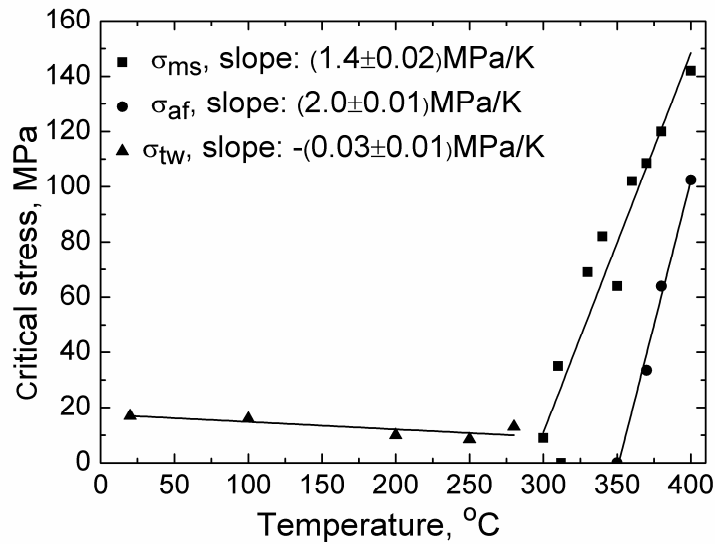

Fig.4. Tensile critical stresses of the forward start, $\sigma_{ms}$ (squares) and reverse finish, $\sigma_{af}$, (circles) martensitic transformations of sample S1 versus temperature. The triangles stand for the detwinning stress in martensitic phase. Linear approximations of the data points are used.

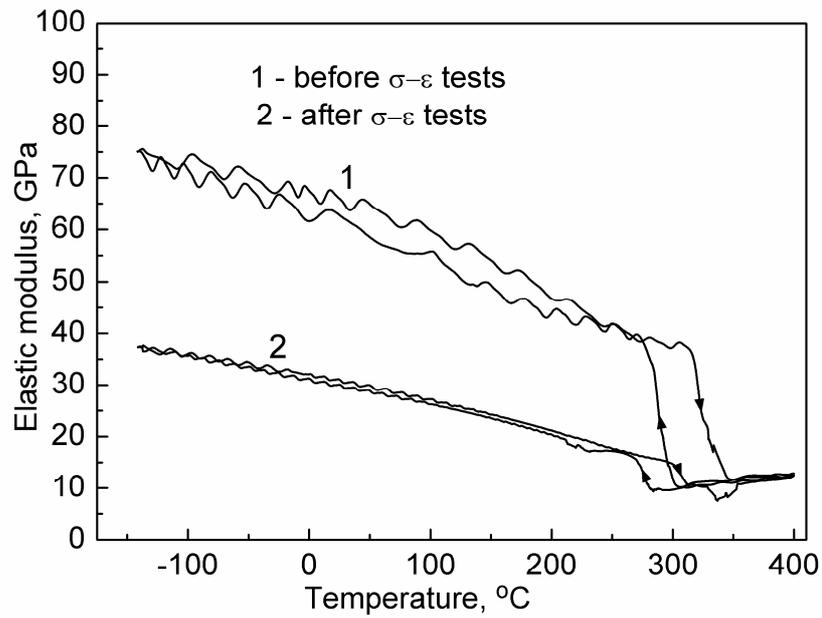

Fig.5. Temperature dependences of the effective elastic modulus of sample S1 in the initial heating-cooling cycle (1) and after thermomechanical cycling according to Route 1 (2). The serration of curves is an instrumental effect.

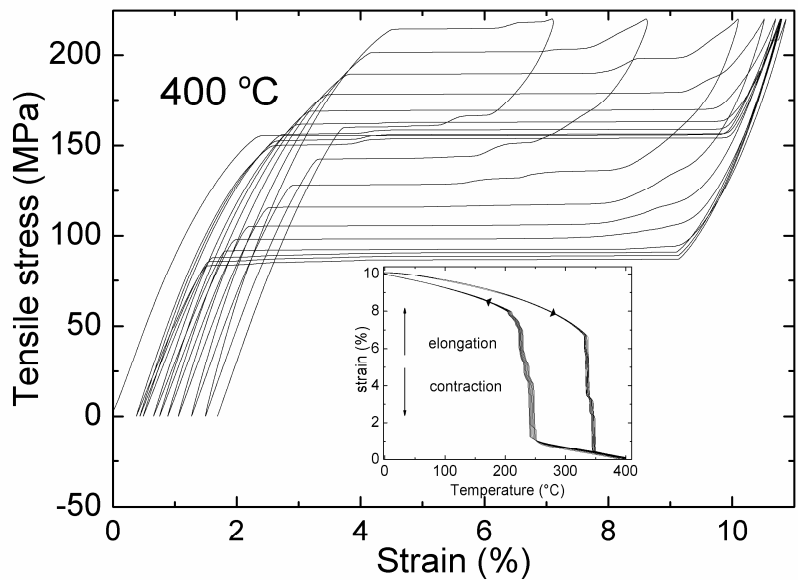

Fig. 6. Ten $\sigma$-$\varepsilon$ superelastic curves of sample S2 recorded at 400°C showing a strong upward shift of the hysteresis loops as a function of cycle number. Inset: ten stress-free temperature dependences of strain measured after cycling shown in Fig.6, revealing a perfect ultra-large TWSME.

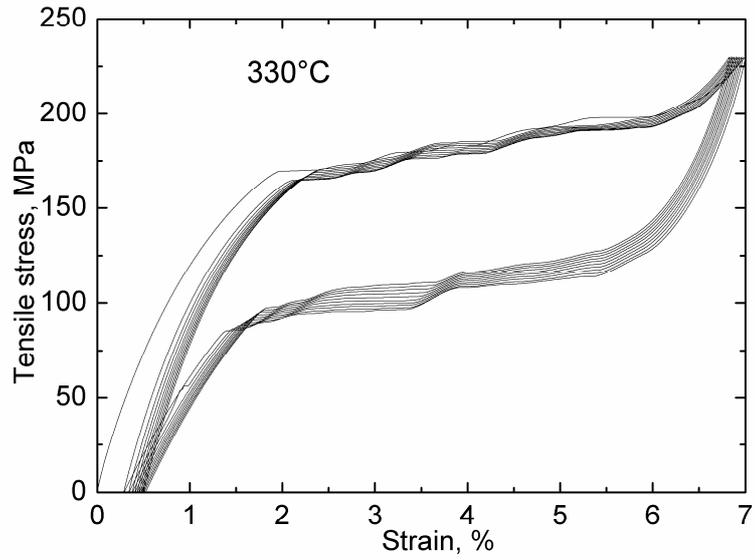

Fig. 7. Ten $\sigma$-$\varepsilon$ tests of sample S2 made at 330°C demonstrating a good cycling stability of the superelastic effect at this temperature.

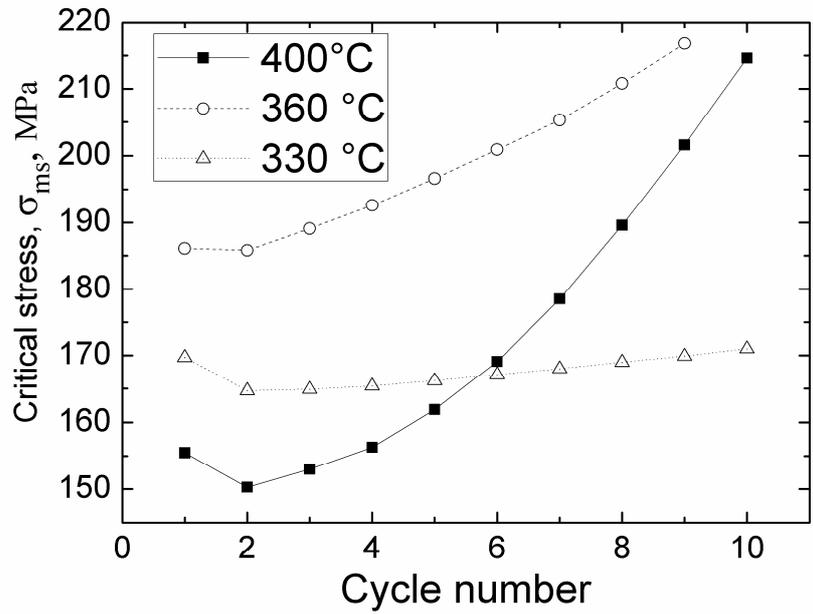

Fig.8. Plots of critical stress as a function of cycle number for sample S2 at different temperatures.

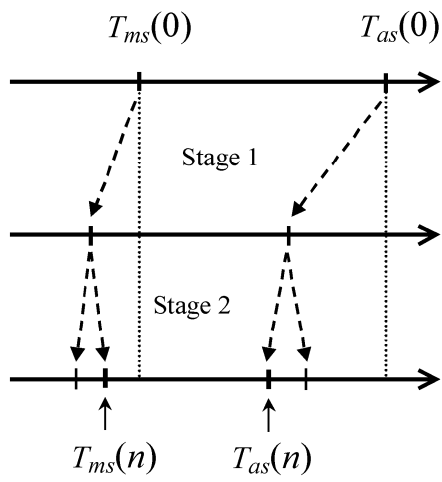

Fig.9. Schematic of the shifting of austenite start and martensite start temperatures under the action of destabilizing internal pressure (Stage 1) and stabilizing internal stress (Stage 2). The solid vertical arrows show the resultant austenite start and martensite start temperatures.